\documentstyle[pra,aps,twocolumn,epsf]{revtex}
\begin{document}

%\twocolumn
\title{Boundary of two mixed Bose-Einstein condensates}

\author{R.A. Barankov}

\address{Department of Physics,
The University of Texas at Austin, Austin, Texas, 78712}

\maketitle

\date{\today}

\begin{abstract}
The boundary of two mixed Bose-Einstein condensates interacting repulsively
was considered in the case of spatial separation at zero temperature.
Analytical expressions for density distribution
of condensates were obtained by solving two coupled nonlinear Gross-Pitaevskii
equations in cases corresponding weak and strong separation. These expressions
allow to consider excitation spectrum of a particle confined in the vicinity
of the boundary as well as surface waves associated with surface tension.
\\

\noindent\verb+PACS Number(s): 03.75.Fi+

\end{abstract}

%\pacs{PACS Number(s): 03.75.Fi}

\section{Introduction}

The experimental realization of Bose Einstein condensation in trapped dilute
gases \cite{Anderson,Davis,Bradley} has allowed to investigate variety of
properties of quantum fluids both theoretically and experimentally. In the
recent years, it has become possible to produce and explore mixtures of
Bose-Einstein condensates corresponding to different internal
states~\cite{Myatt,Hall1,Hall2}.

Theoretical treatment of mixtures \cite{Timmermans,Ao,Pu,Shi} assures that
depending on the relative strength of interactions inside each condensate and
between them it is possible to observe spatial separation.
Experimental realization of such systems~\cite{Myatt,Hall1,Hall2} has given
an opportunity to study both equilibrium properties and dynamics of
separation. Although, structure of the boundary has been qualitatively
analyzed in~\cite{Timmermans,Ao} an analytical distribution of densities
has not been derived yet. Authors of~\cite{Ao} discussed asymptotic behavior
of densities far from the boundary, and on the basis of these estimations
analyzed surface tension.

In the present paper, we explore the boundary between two repulsively
interacting condensates at zero temperature in two limit cases corresponding
weak and strong separation of the condensates. We show that in the case
of weak separation it is possible to derive equations for densities, and using
iteration method solve them analytically. We show that asymptotic behavior
of our solution coincides with those predicted in~\cite{Ao}.
For the sake of completness, we also
provide the solutions of these equations in the case of strong separation.
Importance of the solution lies in the possibility to explore quantitatively
different types of excitations on the boundary.
The structure of the boundary in both cases allows to consider one-particle
excitations as well as surface waves associated with the boundary.
Obtained expressions for dispersion relation of surface waves can be used to
explore that phenomenon experimentally.

The Hamiltonian describing the mixture of two weakly interacting
Bose-Einstein condensates can be written in the form
\begin{equation}\label{ham}
\begin{array}{r}
H=\sum\limits_{i=1,2}\int d{\bf r} \Psi_{i}^{+} \left[-\frac{\hbar^{2}
\nabla^{2}}{2 m_{i}}+V_{i}({\bf r})+
\frac{u_{i}}{2}\Psi_{i}^{+}\Psi_{i}\right]\Psi_{i}\\
+u_{12}\int d{\bf r}\Psi_{1}^{+}\Psi_{2}^{+}\Psi_{1}\Psi_{2}
\end{array}
\end{equation}
Here, $u_{i}=4 \pi \hbar^{2} a_{i}/m_{i}>0$ characterizes the interaction
inside each condensate,
$u_{12}=2 \pi \hbar^{2} a_{12} (m_{1}+m_{2})/(m_{1}m_{2})>0$ -- the
intercondensate interaction, $m_{i}$ -- mass of a particle of each
condensate, $a_{i}$, $a_{12}$ -- corresponding scattering lengths,
$V_{i}(\bf r)$ -- external trapping potentials. Theoretical treatment
of the mixtures \cite{Timmermans,Ao} has shown that
separation takes place when $u_{12}/\sqrt{u_{1} u_{2}}>1$.

Starting with the Hamiltonian~(\ref{ham}) we get Gross-Pitaevskii equations
for condensate wave functions:
\begin{equation}\label{nonsteqs}
\begin{array}{r}
i\hbar\frac{\partial\Psi_{1}}{\partial t}=\left(-\frac{\hbar^{2}
\nabla^{2}}{2 m_{1}}+V_{1}({\bf r})+u_{1}|\Psi_{1}|^{2}+
u_{12}|\Psi_{2}|^{2}\right)\Psi_{1},
\\
i\hbar\frac{\partial\Psi_{2}}{\partial t}=\left(-\frac{\hbar^{2}
\nabla^{2}}{2 m_{2}}+V_{2}({\bf r})+u_{2}|\Psi_{2}|^{2}+
u_{12}|\Psi_{1}|^{2}\right)\Psi_{2}
\end{array}
\end{equation}

As we are interested in studying of stationary solutions of these equations,
then assuming as usual $\Psi_{j} \propto \exp(-i\mu_{j} t)$, where
$\mu_{j}$ -- chemical potentials of the condensates, we obtain two coupled
nonlinear equations for densities of gases
$n_{i}({\bf r})=|\Psi_{i}({\bf r})|^{2}$:
\begin{equation}\label{steqs}
\begin{array}{l}
\mu_{1}=-\frac{\hbar^{2}}{2 m_{1}}\frac{\nabla^{2}\sqrt{n_{1}}}
{\sqrt{n_{1}}}+V_{1}({\bf r})+u_{1}n_{1}+u_{12}n_{2},
\\
\mu_{2}=-\frac{\hbar^{2}}{2 m_{2}}\frac{\nabla^{2}\sqrt{n_{2}}}
{\sqrt{n_{2}}}+V_{2}({\bf r})+u_{2}n_{2}+u_{12}n_{1}
\end{array}
\end{equation}

These equations are essentially nonlinear, so to find solutions we need
to make some simplifications. We assume that $V_{1}({\bf r})=V_{2}({\bf r})$
and consider the case when the size of a boundary between condensates much
less than characteristic length of the trap. In the case of a parabolic trap
potential and Thomas-Fermi regime, it means that $d\ll R_{TF}$, where $d$ is
the size of the boundary, and $R_{TF}$ is the Thomas-Fermi radius of the
atomic cloud. Physically, it helps to avoid the effect of the potential on
the form of the boundary. To simplify calculations further, we also suppose
that separation takes place in one dimension, $z$.
This allows us to write the system of
equations~(\ref{steqs}) in the following form:
\begin{equation}\label{eqs}
\begin{array}{l}
\mu_{1}=-\frac{\hbar^{2}}{2 m_{1}\sqrt{n_{1}}}\frac{d^{2}}{dz^{2}}
\sqrt{n_{1}}+u_{1}n_{1}+u_{12}n_{2},
\\
\mu_{2}=-\frac{\hbar^{2}}{2 m_{2}\sqrt{n_{2}}}\frac{d^{2}}{dz^{2}}
\sqrt{n_{2}}+u_{2}n_{2}+u_{12}n_{1}
\end{array}
\end{equation}

Although, there is no explicit trapping potential in these equations, we
have to impose external conditions on the solutions, that implicitly take
it into account. Let us assume that separation takes place along the $z$
direction, and the condensate with the label ``$1$" is to the right, and
the one with the label ``$2$" is to the left of the boundary. Then,
asymptotically we require:
\begin{equation}\label{conds}
\begin{array}{l}
n_{1}(z\to +\infty)\to n_{10},\, n_{1}(z\to -\infty)\to 0,\\
n_{2}(z\to -\infty)\to n_{20},\,n_{2}(z\to +\infty)\to 0,
\end{array}
\end{equation}
where $n_{10}$, $n_{20}$ -- equilibrium densities of condensates far from
the boundary. Substituting these conditions in equations~(\ref{eqs}) for
densities, and using the obvious condition of equilibrium we obtain
\begin{equation}\label{prconds}
\begin{array}{l}
\mu_{1}=u_{1}n_{10},\,\mu_{2}=u_{2}n_{20},\\
P_{1}=u_{1}n_{10}^{2}/2=P_{2}=u_{2}n_{20}^{2}/2
\end{array}
\end{equation}
Here, we used well-known expression for the pressure of a homogeneous
weakly-interacting Bose gas.
That condition connects equilibrium densities of condensates far from
the boundary.

To reduce the number of parameters in equations~(\ref{eqs}) we notice
that it is possible to exclude difference of masses by the change:
\begin{equation}\label{mass}
\begin{array}{l}
u_{1}^{*}=u_{1}m_{1}/m_{2},\,u_{2}^{*}=u_{2}m_{2}/m_{1},\\
n_{1}^{*}=n_{1}\sqrt{m_{2}/m_{1}},\,n_{2}^{*}=n_{2}\sqrt{m_{1}/m_{2}},\\
\mu_{1}^{*}=u_{1}^{*}n_{10}^{*},\,\mu_{2}^{*}=u_{2}^{*}n_{20}^{*},\\
m^{*}=\sqrt{m_{1}m_{2}}
\end{array}
\end{equation}

We get the same equations~(\ref{eqs}) and conditions~(\ref{conds}),
(\ref{prconds}) but for quantities with stars and with the same mass
$m^{*}$. To simplify notations, in the following discussion we omit stars.
The generalization for different masses can be easily done by following
the above rules.

We can solve equations for densities~(\ref{eqs}) with
conditions~(\ref{conds}),(\ref{prconds}) analytically in two limit
cases for weak separation
when $\Delta=u_{12}/\sqrt{u_{1} u_{2}}-1\ll 1$, and strong separation
when $\Delta\gg 1$. Notice, that both parameters $u_{12}$ and $\Delta$
are not affected by the above procedure of mass difference excluding.

\section{Weak separation}
Let us consider weak separation when condition
$\Delta=u_{12}/\sqrt{u_{1} u_{2}}-1\ll 1$ is satisfied. We expect that
in the simplest case when $u_{1}=u_{2}$ the whole density of a gas
$n(z)=n_{1}(z)+n_{2}(z)$ is approximately constant. That is why, to find
the solution it is natural instead of $n_{1}$, $n_{2}$ to introduce
other quantities and solve equations~(\ref{eqs}) using small parameter
$\Delta$. Consider the functions:
\begin{equation}\label{new}
\begin{array}{l}
\rho=(u_{1}/u_{2})^{1/4}n_{1}+(u_{2}/u_{1})^{1/4}n_{2},\\
g=\left[(u_{1}/u_{2})^{1/4}n_{1}-(u_{2}/u_{1})^{1/4}n_{2}\right]/\rho
\end{array}
\end{equation}

Conditions~(\ref{conds}), (\ref{prconds}) give us simple asymptotic behavior
of these functions:
\begin{equation}\label{condsnew}
\begin{array}{l}
\rho(z\to\pm\infty)\to\rho_{0}=\sqrt{n_{10}n_{20}},\\
g(z\to\pm\infty)\to\pm 1
\end{array}
\end{equation}

Densities of condensates ``1" and ``2" are easily obtained if functions
$\rho$ and $g$ are known:
\begin{equation}
\begin{array}{l}
n_{1}=(u_{2}/u_{1})^{1/4}\rho[1+g]/2,\\
n_{2}=(u_{1}/u_{2})^{1/4}\rho[1-g]/2
\end{array}
\end{equation}

It is straightforward to derive equations for $\rho$ and $g$:
\begin{equation}\label{eqsnew}
\begin{array}{r}
\frac{\sqrt{\rho}^{\,\prime\prime}}{\sqrt{\rho}}-
\frac{\rho}{\rho_{0}}\left[1+\alpha g +\frac{\Delta}{2}
(1-g^{2})\right]\\
=\frac{g^{\prime\,2}}{4(1-g^{2})}-1-\alpha g, \\
\frac{g^{\prime\prime}}{1-g^{2}}+\frac{2\sqrt{\rho}^{\,\prime}
g^{\prime}}{\sqrt{\rho}(1-g^{2})}+\frac{g g^{\prime\,2}}
{(1-g^{2})^{2}}+2\alpha\\
=\frac{\rho}{\rho_{0}}\left[2\alpha+\Delta(\alpha-g)\right]\\
\end{array}
\end{equation}
Here, $f^{\prime}=\xi_{0}\frac{d f}{d z}$, and we also introduced
$\alpha=\frac{\sqrt{u_{1}}-\sqrt{u_{2}}}{\sqrt{u_{1}}+\sqrt{u_{2}}}$ and
$1/\xi_{0}^{2}=m(u_{1}u_{2})^{1/4}(\sqrt{u_{1}}+\sqrt{u_{2}})
\rho_{0}/\hbar^{2}$.

To find asymptotic solutions of equations~(\ref{eqsnew}) in the case of
$\Delta\ll 1$ we use iteration method. Let us suppose that terms
with derivatives of $\rho$ are much smaller than the others. We will
justify this assumption in the end of calculations.

Neglecting terms with derivatives of $\rho$, from equations~(\ref{new})
we get
\begin{equation}
\begin{array}{l}
\rho/\rho_{0}=1-\frac{g^{\prime\,2}+2\Delta(1-g^{2})^{2}}{4(1-g^{2})
(1+\alpha g)},\\
\frac{g^{\prime\prime}}{1-g^{2}}+\frac{2g+\alpha(1+ g^{2})}{2(1-g^{2})
(1+\alpha g)}g^{\prime\,2}+\frac{\Delta (1-\alpha^{2})g}{1+\alpha g}=0
\end{array}
\end{equation}

The equation for $g$ can be solved by substitution $g^{\prime}=f$, so taking
into account conditions~(\ref{condsnew}) we get that $g$ is the solution of
the equation
\begin{equation}\label{eqsg}
g^{\prime}=\sqrt{\Delta(1-\alpha^{2})}\frac{(1-g^{2})}{\sqrt{1+\alpha g}}
\end{equation}

At this point it is clear that assumptions we made were correct. Namely, we
obtain that $\sqrt{\rho}^{\,\prime}\propto \Delta^{3/2}$,
$\sqrt{\rho}^{\,\prime\prime}\propto \Delta^{2}$,
$g^{\prime}\propto\sqrt{\Delta}$, so neglected terms by $\Delta$ times less
than the others.

Now we can write down solutions for both $\rho$ and $g$ in the parametric
form
\begin{equation}\label{solutions}
\begin{array}{r}
1-\frac{\rho}{\rho_{0}}=\frac{\Delta}{4}\frac{1-g^{2}}{(1+\alpha g)^{2}}
[3-\alpha^{2}+2\alpha g],\\
\frac{z-z_{0}}{\xi_{0}/\sqrt{\Delta(1-\alpha^{2})}}=\frac{\sqrt{1+\alpha}}{2}
\,\ln|\frac{\sqrt{1+\alpha g}+\sqrt{1+\alpha}}{\sqrt{1+\alpha g}-
\sqrt{1+\alpha}}|\\
-\frac{\sqrt{1-\alpha}}{2}\ln|\frac{\sqrt{1+\alpha g}+\sqrt{1-\alpha}}
{\sqrt{1+\alpha g}-\sqrt{1-\alpha}}|
\end{array}
\end{equation}
Here, the second equation is the solution of~(\ref{eqsg}). In these
expressions an arbitrary constant $z_{0}$ defines the position of the
boundary, and in the case of finite geometry with given average number
of particles in condensates can be obtained from the condition of equal
pressures~(\ref{prconds}). There is also physically obvious symmetry
$z-z_{0}\to z_{0}-z$, $g\to-g$, $\alpha\to-\alpha$.

\begin{figure}
\begin{center}
\epsfxsize=3in
\epsffile{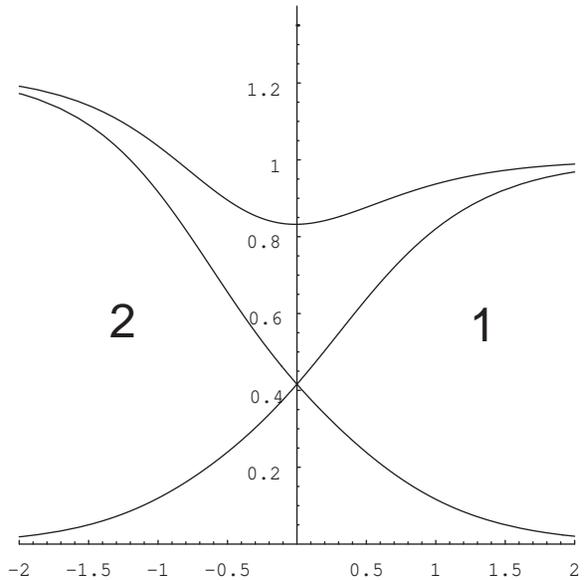}
\caption{
Distribution of densities in the case of weak separation.
$\Delta=0.3$, $\alpha=0.1$. Densities are in
the scale of $n_{10}$.Distance is in the scale of
$\xi_{0}/\sqrt{\Delta(1-\alpha^{2})}$. There is a well in the whole density
profile.
}
\label{fig:well}
\end{center}
\end{figure}

\begin{figure}
\begin{center}
\epsfxsize=3in
\epsffile{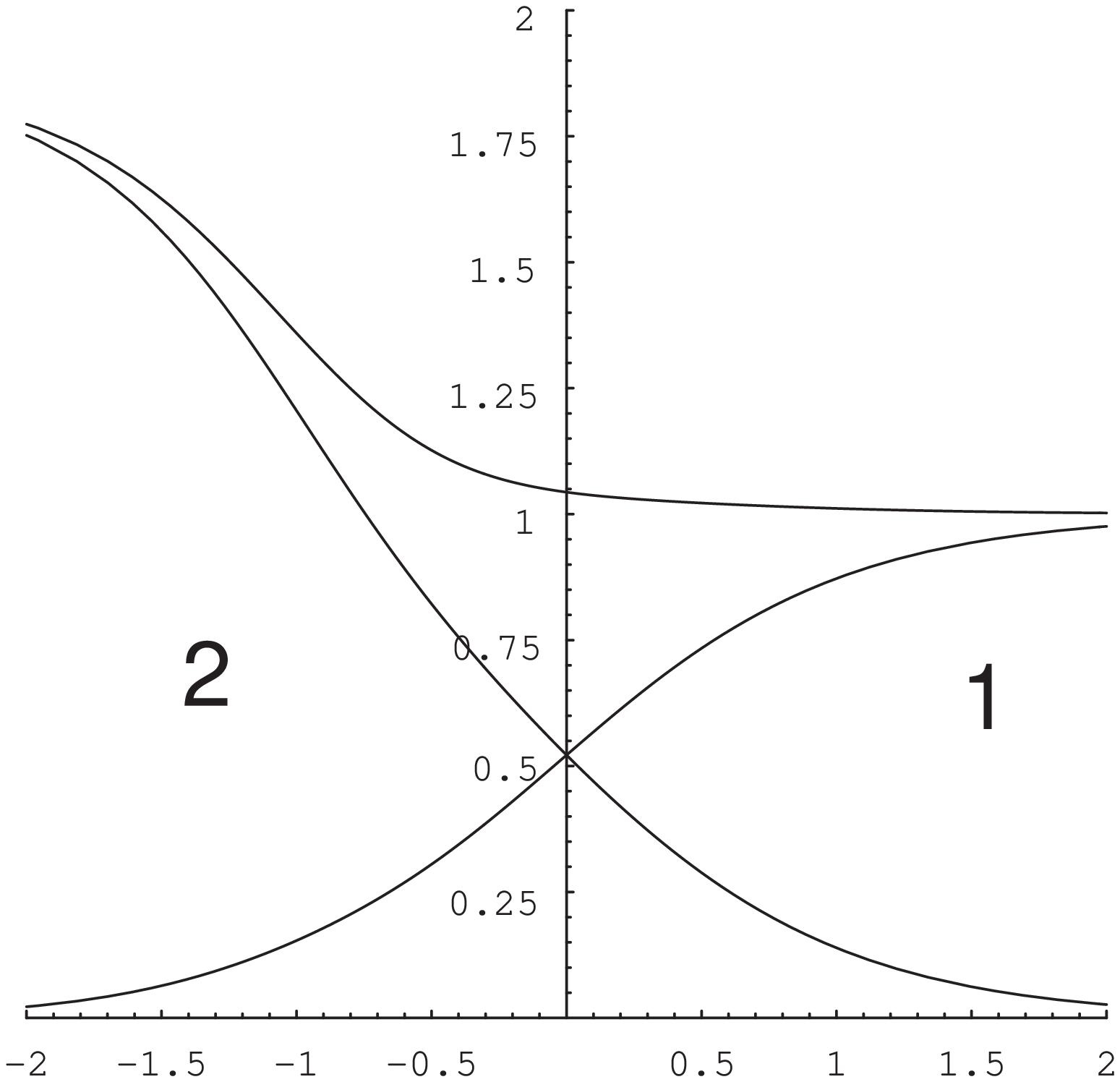}
\caption{
Distribution of densities in the case of weak separation.
$\Delta=0.3$, $\alpha=0.3$. Densities are in
the scale of $n_{10}$.
Distance is in the scale of
$\xi_{0}/\sqrt{\Delta(1-\alpha^{2})}$
}
\label{fig:nowell}
\end{center}
\end{figure}

Finally, the densities of condensates have the following form:
\begin{equation}\label{dens}
\begin{array}{l}
n_{1}=\frac{n_{10}}{2}\left(1-\frac{\Delta}{4}\frac{1-g^{2}}
{(1+\alpha g)^{2}}[3-\alpha^{2}+2\alpha g]\right)\left[1+g\right],\\
n_{2}=\frac{n_{20}}{2}\left(1-\frac{\Delta}{4}\frac{1-g^{2}}
{(1+\alpha g)^{2}}[3-\alpha^{2}+2\alpha g]\right)\left[1-g\right]
\end{array}
\end{equation}
With the parametric equation for $g$, these densities are the main result
of the paper. On fig With the use of the desribed method, it is possible to derive
expressions for densities in the next orders of the small parameter $\Delta$
in the form of asymptotic series. As follows from the above estimations
of neglected terms in equations~(\ref{eqsnew}) the next order
is proportional to $\Delta^{2}$.  The typical dependence of densities on
the distance from the boundary is shown in Fig.~\ref{fig:well} and
Fig.~\ref{fig:nowell}.

It is interesting to notice that the whole density $n=n_{1}+n_{2}$ at some
values of parameters has a well on the boundary as shown in Fig.~\ref{fig:well}.
In the case of weak
separation we see that it is broad and shallow. This result is a consequence
of the interparticle interactions.
The interparticle potential for each condensate acts in some sense as a wall,
so we expect that probability of a particle to be close to the boundary
decreases, which means lower density.
The existence of such a well allows to consider the possibility of confining
of a particle of another sort in the vicinity of the boundary, which is
discussed below.

Although, we can not further simplify solutions~(\ref{solutions}), it is
interesting to derive asymptotic behavior of these functions in particular
limits:
\begin{equation}\label{asymp}
\begin{array}{l}
1-g(z\to +\infty)\propto \exp[-\frac{2 z \sqrt{\Delta}}{\xi_{2}}],\\
1+g(z\to -\infty)\propto \exp[\frac{2 z \sqrt{\Delta}}{\xi_{1}}],\\
g(z\to z_{0})\to \frac{(z-z_{0})\sqrt{\Delta(1-\alpha^{2})}}{\xi_{0}}
\end{array}
\end{equation}
Here, $\xi_{i}=\hbar/\sqrt{2m_{i}\mu_{i}}$ -- correlation lengths of
condensates, defined by chemical potential $\mu_{1}$ and $\mu_{2}$, and
we used~(\ref{mass}) to include mass difference.
The asymptotic behavior far from the boundary of each condensate is defined
by its correlation length. As we see, the size of the boundary can be
approximated as $d\sim (\xi_{1}+\xi_{2})/\sqrt{\Delta}$ and for
$\Delta\ll 1$ appears to be much larger than correlation lengths.

Solutions have the simplest form when $\alpha=0$:
\begin{equation}\label{simdens}
\begin{array}{l}
n_{1}(z)=\frac{n_{0}}{2}\left(1-\frac{3\Delta}{4 \cosh^{2}
[\frac{z\sqrt{\Delta}}{\xi_{0}}]}\right)
\left(1+\tanh[\frac{z\sqrt{\Delta}}{\xi_{0}}]\right),\\
n_{2}(z)=\frac{n_{0}}{2}\left(1-\frac{3\Delta}{4 \cosh^{2}
[\frac{z\sqrt{\Delta}}{\xi_{0}}]}\right)
\left(1-\tanh[\frac{z\sqrt{\Delta}}{\xi_{0}}]\right)\\
\end{array}
\end{equation}
Here, we choose $z_{0}=0$.

\section{Strong separation}
To analyze the case of strong separation
$\Delta=u_{12}/\sqrt{u_{1}u_{2}}\gg 1$ (we use the same notation but for
another quantity) we start from density equations~(\ref{eqs}). In this case
we expect that density on the boundary will be approximately zero because
interparticle interactions make it almost impossible for one condensate to
penetrate inside the other. To estimate the
density of condensates on the boundary we can use the fact that second
derivatives of wave functions should be approximately zero there. That makes
system~(\ref{eqs}) a set of two linear equations with solution:
\begin{equation}
\begin{array}{l}
n_{1B}=\frac{n_{10}}{\Delta+1}\approx\frac{n_{10}}{\Delta}\ll n_{10},\\
n_{2B}=\frac{n_{20}}{\Delta+1}\approx\frac{n_{20}}{\Delta}\ll n_{20}
\end{array}
\end{equation}

This allows us in the zero approximation to use simple conditions for the
densities $n_{1}(z\le0)=n_{2}(z\ge0)=0$. Then equations~(\ref{eqs}) have
simple form:
\begin{equation}
\begin{array}{l}
\mu_{1}=-\frac{\hbar^{2}}{2 m_{1}\sqrt{n_{1}}}\frac{d^{2}}{dz^{2}}
\sqrt{n_{1}}+u_{1}n_{1},\,\mbox{ for } z\ge0,\\
\mu_{2}=-\frac{\hbar^{2}}{2 m_{2}\sqrt{n_{2}}}\frac{d^{2}}{dz^{2}}
\sqrt{n_{2}}+u_{2}n_{2},\,\mbox{ for } z\le0
\end{array}
\end{equation}

Solutions are easily obtained:
\begin{equation}\label{strong}
\begin{array}{l}
n_{1}(z\ge 0)=n_{10}\tanh^{2}[\frac{z}{\sqrt{2}\xi_{1}}],\\
n_{2}(z\le 0)=n_{20}\tanh^{2}[\frac{z}{\sqrt{2}\xi_{2}}]
\end{array}
\end{equation}
Here, $n_{10}$ and $n_{20}$ are connected by the condition of equal
pressures~(\ref{prconds}), and we choose the position of the boundary
at $z=0$.

\begin{figure}
\begin{center}
\epsfxsize=3in
\epsffile{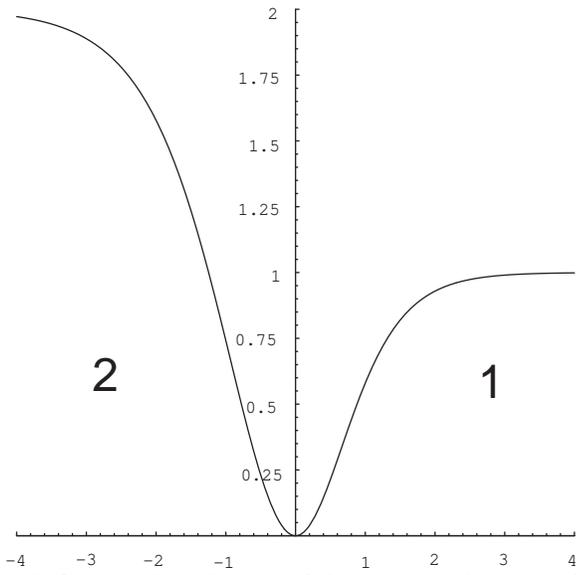}
\caption{
Distribution of densities in the case of strong separation when  $u_{1}/u_{2}=4$.
Densities are in
the scale of $n_{10}$.
Distance is in the scale of $\sqrt{2}\xi_{1}$.
}
\label{fig:strong}
\end{center}
\end{figure}

The size of the boundary in this case is approximately
$d\approx 2\sqrt{2}(\xi_{1}+\xi_{2})$. The dependence of the densities on the distance
from the boundary is shown in Fig.~\ref{fig:strong}.
As in the previous section, we see that there
is again a well in the whole density but in the case of strong separation it
becomes narrower and deeper in comparison with the one for weak separation.

\section{One-particle excitations on the boundary}
The existence of a well in the whole density allows to consider confining of a
particle of another sort in the vicinity of the boundary. As a general
property of a quantum-mechanical motion in a one-dimensional well there
always exists such a confined state. As an example, we consider the simplest
case when $\alpha=0$ and a particle of another sort interacts with both
condensates repulsively with the same constant $\lambda$.
The Schr\"odinger equation for the wave function of a particle with the mass
$M$ has the form:
\begin{equation}\label{part}
\left[-\frac{\hbar^{2}}{2 M}\frac{d^{2}}{d\,z^{2}}+\lambda n(z)\right]
\phi=E \phi
\end{equation}

We can solve equation~(\ref{part}) for weak and strong separation cases
simultaneously. It has universal form
\begin{equation}
\frac{d^{2}\phi}{d\,z^{2}}+\frac{2 M}{\hbar^{2}}\left(\epsilon+
\frac{U_{0}}{\cosh^{2}[\beta z]}\right)\phi=0,
\end{equation}
where $\epsilon=E-\lambda n_{0}$, and $U_{0}=3\Delta\lambda n_{0}/4$,
$\beta=\sqrt{\Delta}/\xi_{0}$ for weak; and $U_{0}=\lambda n_{0}$,
$\beta=1/(\sqrt{2}\xi_{0})$ for strong separation. The spectrum of energy
$\epsilon$ is well-known:
\begin{equation}
\begin{array}{l}
\epsilon_{j}=-\mu\frac{\Delta m}{4M}\left[-(1+2j)+\sqrt{1+3
\frac{M \lambda}{m u}}\right]^{2} \,\mbox{``weak"},\\
\epsilon_{j}=-\mu\frac{m}{8M}\left[-(1+2j)+\sqrt{1+8
\frac{M \lambda}{m u}}\right]^{2} \,\mbox{``strong"}
\end{array}
\end{equation}
where $\mu=un_{0}$ -- chemical potential, $j=0,1,\dots$ and the condition
that an expression in $[\dots]$ is positive defines the upper limit for $j$.
There is always at least one state with $j=0$.

\section{Surface waves on the boundary}
There is also another type of excitations associated with the boundary.
As we see condensate density distributions give rise to nonzero surface
tension which was previously analyzed in~\cite{Ao}, and we use the
expression for surface tension derived there:
\begin{equation}\label{gentension}
\sigma=\frac{1}{2}\int\limits_{-\infty}^{+\infty}dz\sum\limits_{i=1,2}
\frac{\hbar^{2}}{2 m_{i}} \left(\frac{d\sqrt{n_{i}}}{dz}\right)^{2}
\end{equation}

Substituting expressions~(\ref{dens}) for densities in the case of weak
separation and taking into account only first nonzero order in $\Delta$
we obtain for surface tension
\begin{equation}\label{tensionweak}
\sigma_{w}=\frac{P\sqrt{\Delta}}{6}\left[
\begin{array}{r}
(\xi_{1}+\xi_{2})\frac{2\alpha^{2}-1+\sqrt{1-\alpha^{2}}}
{\alpha^{2}}\\
-(\xi_{1}-\xi_{2})\frac{1-\sqrt{1-\alpha^{2}}}{\alpha}
\end{array}
\right],
\end{equation}
where $\alpha=\left(m_{1}\sqrt{u_{1}}-m_{2}\sqrt{u_{2}}\right)/
\left(m_{1}\sqrt{u_{1}}+m_{2}\sqrt{u_{2}}\right)$ and we used~(\ref{mass})
to include mass difference; $\xi_{1}$, $\xi_{2}$ -- correlation lengths
of condensates; $P$ -- the pressure given by~(\ref{prconds}).
When $\alpha=0$, $\sigma_{w}=P\sqrt{\Delta}(\xi_{1}+\xi_{2})/4$

In the case of strong separation we use expressions~(\ref{strong}) to get
surface tension
\begin{equation}\label{tensionstrong}
\sigma_{s}=\frac{P\sqrt{2}}{3}(\xi_{1}+\xi_{2})
\end{equation}

Let us notice that expressions~(\ref{tensionweak}) and~(\ref{tensionstrong})
differ from the ones obtained in~\cite{Ao}. Although, in qualitative sense our
expressions coincide with those of~\cite{Ao}, using our method of solution
we can get a general expression applicable for variety of parameters, and for
example retrieve the correct numerical factor for the case considered
in~\cite{Ao}. As follows from the general expression for surface
tension~(\ref{gentension}), we need to know the behavior of densities
not only far but also in the vicinity of the boundary. That is why, estimate
character of expressions for densities in~\cite{Ao} could only give
qualitative answer for surface tension.

For velocities smaller than the speed of sound we can consider gas as
incompressible, so it is possible to write down the usual hydrodynamic
equations and find dispersion relation of surface waves on the
boundary associated with surface tension. Suppose that we have condensates
in a ``box" and the position of the boundary is defined by
the distances $L_{1}$ and $L_{2}$ from the ``box" walls, where labels ``1",
``2" correspond to the side with the same condensate, and the box sizes for
other directions are $L_{x}$, $L_{y}$.
Taking into account the fact that velocities are zero on the walls we get
the dispersion relation
\begin{equation}\label{wave}
\omega^{2}(k)=\frac{\sigma\,k^{3}}{\rho_{1}\coth(k L_{1})+
\rho_{2}\coth(k L_{2})},
\end{equation}
where $\sigma$ is the surface tension; $\rho_{10}=m_{1}n_{10}$,
$\rho_{20}=m_{2}n_{20}$ -- mass densities; wave vector
along the boundary
$k=\sqrt{(\pi n_{x}/L_{x})^{2}+(\pi n_{y}/L_{y})^{2}}$,
$n_{x}, n_{y}=0,1\dots$, not $n_{x}=n_{y}=0$;
$L_{1}$, $L_{2}$ -- sizes of condensates in the direction perpendicular to
the boundary. We assume that $L_{1}$, $L_{2}$ are much larger than the size
of the boundary $d$.

For dispersion relation we consider two limit cases corresponding to
wavelengths $\lambda\ll L_{i}$ ($k L_{i}\gg 1$) and
$\lambda\gg L_{i}$ ($k L_{i}\ll 1$). Let us notice that the second case
is possible only if $L_{i}\ll L_{x,y}$.

In the first case we obtain
\begin{equation}
\omega(k)=\left(\frac{\sigma}{\rho_{10}+\rho_{20}}\right)^{1/2}k^{3/2}
\end{equation}

In the long wavelength limit we get
\begin{equation}
\omega(k)=\left(\frac{\sigma L_{1} L_{2}}{\rho_{10}L_{2}+\rho_{20}L_{1}}
\right)^{1/2}k^{2}
\end{equation}

As we see, for weak separation $\omega\propto \Delta^{1/4}$, and surface
waves are relatively ``soft" in that case, which enables to consider them
as a dissipative channel for other condensate excitations.

\section{Conclusion}
We performed the analysis of the boundary of two Bose-Einstein condensates
interacting repulsively in the limit cases corresponding weak and strong
separation at zero temperature.

For weak separation we obtained solutions~(\ref{dens}) of two coupled
nonlinear Gross-Pitaevskii equations using small parameter $\Delta$.
The solutions show that the penetration depth of the condensate ``$i$"
inside the other is estimated as $\xi_{i}/\sqrt{\Delta}$, so that the size
of the boundary $d\sim (\xi_{1}+\xi_{2})/\sqrt{\Delta}$ is much larger than
correlation lengths, which was obtained experimentally~\cite{Hall1}.
There is also a well in the full density profile, which
is the consequence of the wave function behavior near the boundary.
In general, the proposed method of obtaining density distributions for the
weak separation case if desired can be expanded to obtain expressions for
the next orders of $\Delta$.

We also considered the case of strong condensate separation but restricted
ourselves to the zero order approximation. In that case the size of
the boundary is $d\sim (\xi_{1}+\xi_{2})$, and the whole density of gases
goes approximately to zero on the boundary.

The existence of the well in the density profile at some parameters allowed
us to consider one-particle excitations on the boundary. Using the expressions
for densities we found excitation spectrum of a particle in the simplest case
when constants of interaction of the particle with condensates are the same,
and distribution of densities on the boundary corresponds to $\alpha=0$.
The generalization to other cases can be easily done with the use of
expressions~(\ref{dens}),(\ref{strong}) for density distributions. Let
us notice that the existence of a potential well depends on the interaction
constants of a particle with condensates as well as the relation between
interaction constants of the condensates. Observation of such confined
states can be possible only if temperature are smaller than the potential
well.

It was shown that there also exist collective excitations associated with
surface tension. The expressions for surface tension were obtained for both
weak and strong separation. The dispersion relation for surface waves was
analysed in the case where condensates fill finite volumes. The dispersion
relation has different forms in two cases corresponding to short- and
long-wavelength limit. In the case of weak separation ``soft" surface modes
can be a dissipative channel of other condensate excitations.

Author gratefully acknowledges valuable discussions with Yu.M. Kagan and
other members of the Theoretical Division of RRC ``Kurchatov Institute",
Russia.

\end{document}